\documentclass[preprint,aps,prd,superscriptaddress,groupedaddress,showpacs,nofootinbib]{revtex4-1}
\usepackage{amsmath}
\usepackage{graphicx}
\usepackage{dcolumn}
\usepackage{bm}
\usepackage{amssymb}

\def\npb#1#2#3{Nucl. Phys. B #1 (#3) #2}

\def\plb#1#2#3{Phys. Lett. B #1 (#3) #2}

\def\phr#1#2#3{Phys. Rep. #1 (#3) #2}

\newcommand{\de}{\mathrm{d}}
\newcommand{\be}{\begin{eqnarray}}
\newcommand{\ee}{\end{eqnarray}}
\newcommand{\statfactor}{\frac{4 \pi g_f g_s }{(2 \pi)^3}}
\newcommand{\statfactornogs}{\frac{4 \pi g_f }{(2 \pi)^3}}
\usepackage{graphicx}
\newcommand{\ave}[1]{\left\langle #1 \right\rangle}
\newcommand{\order}[1]{ \mathcal{O} \left( #1 \right) }

\newcommand{\eqcomma}{\phantom{A},\phantom{A}}

  \newcommand{\QCD}{\mbox{\tiny{QCD}}}
  \newcommand{\lqcde}[1]{\Lambda_{\QCD}^{#1}}
  \newcommand{\Lqcd}{\Lambda_{\QCD}}
  \newcommand{\lqcd}{\Lambda_{\QCD}}
\newcommand{\eq}{\begin{equation}}
\newcommand{\qe}{\end{equation}}
\newcommand{\eqa}{\begin{eqnarray}}
\newcommand{\qea}{\end{eqnarray}}

\begin{document}

\title{Quarkyonic Percolation and deconfinement at finite density and number of colors}

\author{Stefano Lottini}

\affiliation{
ITP,  J.W.~Goethe Universit\"at, Max-von-Laue-Stra\ss{}e 1, 60438 Frankfurt am Main, Germany}
\author{Giorgio Torrieri}

\affiliation{
FIAS, J.W.~Goethe Universit\"at, Ruth-Moufang-Stra\ss{}e 1, \qquad 60438 Frankfurt am Main, Germany and \\ Pupin physics laboratory, Columbia University, New York, NY10027, USA
}
\begin{abstract}
	We examine the interplay between the percolation and the deconfinement phase transitions of Yang-Mills
	matter at finite temperature, quark chemical potential $\mu_Q$ and number of colors $N_c$.
	We find that, whereas the critical $N_c$ for percolation goes down with density, the critical $N_c$ for confinement generally goes up.
	Because of this, Yang-Mills matter falls into two qualitatively different regimes: the ``low-$N_c$ limit'',
	where percolation does not occur because matter deconfines before it percolates, and the ``high-$N_c$ limit'',
	where there are three distinct phases characterizing Yang-Mills matter at finite temperature and density: confined, deconfined and confined but percolating matter.
	The latter can be thought of as the recently conjectured ``quarkyonic phase''.
	We attempt an estimate of the critical $N_c$, to see if the percolating phase can occur in our world.
	We find that, while percolation will not occur at normal nuclear density as in the large-$N_c$ limit,
	a sliver of the phase diagram in $N_c$, energy density and baryonic density where percolation occurs
	while confinement persists is possible.
	We conclude by speculating on the phenomenological properties of such percolating ``quarkyonic'' matter,
	and suggest avenues to study it quantitatively and to look for it in experiment.
\end{abstract}

\maketitle
\section{Introduction: The phase diagram at large $N_c$}
\label{sec:intro}
The ``large number of colors'' approach \cite{thooft,witten} has been a promising way to simplify some of
the tremendous mathematical difficulties inherent in handling non-perturbative features of Yang-Mills theory.
The idea is to take the number of colors $N_c$ to infinity while taking the Yang-Mills coupling constant $g_{YM}$
to zero in such a way that $g_{YM}^2 N_c=\lambda$ stays constant, defined at some perturbative fixed scale.
Numerical results, obtainable by plugging in $N_c=3$, should be correct within $\frac{1}{N_c} \sim 30 \%$ or so,
and hence this simplified theory should be enough for a qualitative estimate.

While this theory shares with QCD its non-perturbative nature (strong coupling arises at a scale $\sim N_c^0$,
parametrically similar therefore to the QCD scale of $\lqcd \simeq 250$ MeV),
this approach has led to some important qualitative results:
the fact that in a confined regime mesons are quasi-particles \cite{thooft} while baryons are classical states
\cite{witten} can be explained in this large-$N_c$ limit. Features of QCD such as the dominance
of planar diagrams (and hence the string description of gluon propagators and extension into the gauge/string correspondence \cite{maldacena})
and the OZI rule are also well explained with $N_c$-counting.
This has made large $N_c$ a useful tool for phenomenological as well as theoretical analysis \cite{manoharrev}.

The large-$N_c$ limit, however, has some qualitative differences from physical QCD too,differences too big to be put down as a $30 \%$ correction:

Due to the identification of confinement with center symmetry restoration \cite{pol0,pol1,conf1},
deconfinement is a first order phase transition in the large-$N_c$ limit provided the number of light flavors $N_f \ll N_c$;
it is a smooth crossover in our $N_c=3$ world \cite{latt1,latt2}.

Nuclear matter is a tightly bound crystal in the large-$N_c$ limit \cite{crystal}, whereas it is a liquid in our world
\cite{liq0,liq1,liq2,liq3,liq4,liq5,liq6,liq7,liqlat}.
The latter feature is a consequence of the fact that in the large-$N_c$ limit the inter-baryon binding energy scales as the baryon mass, $N_c \Lqcd$.
In reality, the scale of inter-nuclear forces is around $\sim \mathcal{O}\big(\Lqcd/(10$---$100)\big)$,
a ``hierarchy problem'' which, given the soundness of the large-$N_c$ description, needs to be resolved.

Given the considerations above, a {\em phase transition} in $N_c$, between $N_c=3$ and $N_c \rightarrow \infty$,
is a plausible resolution of some of these issues \cite{vdw,percolation1,phase0}.
The existence of two possibly linked transitions
in $N_c$ is in fact fairly certain, due to the arguments above: if we could keep $N_f$ constant
($N_f \geq 1$ for baryons to exist) and increase $N_c$, we would find a critical point for confinement at zero quark chemical potential
(somewhere between the cross-over in our world and a first order transition at $N_f/N_c \rightarrow 0$ \cite{panero1,panero2})
and a liquid-crystal transition for matter at high chemical potential
(since the large-$N_c$ matter is crystalline \cite{crystal}, and a crystal-liquid transition is
typically associated to a phase transition due to a change in translational symmetries).

As discussed in \cite{vdw} (and hinted at from gauge/string calculations \cite{lippert,rozali}), the crystal-liquid transition
is linked to the nuclear matter hierarchy problem: the classical picture of the baryon necessarily entails an $N_c$
much larger than $N_N\sim\order{10}$, where $N_N$ is the number of neighbors in a densely packed system.
Below this limit, one can not ignore the Pauli exclusion principle in the {\em non-color part} of the baryonic wavefunction.
This raises the energy cost of compressing baryonic matter by $\sim N_c \lqcde{3}$,
and hence most likely lowers the equilibrium density to values lower than $\sim \lqcde{3}$ (in fact, even $\ll m_\pi^3$).
Since inter-quark interactions are suppressed by the confinement scale, and pionic exchanges are $\sim e^{-r m_\pi}$,
the nuclear forces get weaker, hence the critical point of the nuclear liquid-gas phase transition happens at $T,\mu_Q \ll \Lqcd$.

Going further is hampered by the fact that our understanding of corrections to $N_c$ in this regime is limited.
The regime in which this transition occurs is inevitably strongly coupled, making perturbative calculations untenable.
Standard methods of lattice QCD can not be used, since $\mu_Q/T \sim \order{1}$ \cite{latmu1,latmu2,latmu3}.
And, as discussed in \cite{nicolini}, gauge/gravity techniques are also unreliable since this transition,
by its very nature, is {\em quantum-gravitational}, something of which we have a very limited understanding.

The only possible way to move forward, then, is to investigate models which are simple and qualitative, yet are universally applicable.
One suggested way to describe phase transitions in Yang-Mills is via percolation \cite{perc1,perc2,perc3}.
The idea is that, at increasing energies, the increasing parton densities will make partons of different hadrons ``overlap''
as their interaction cross-section becomes of the order of inter-parton spacing.
It is logical to associate this transition to  deconfinement, where a quark can propagate throughout the hot medium
rather than being confined to the hadron size, whose natural scale is $\sim \lqcde{-1} \sim 1$ fm.
While these analogies might touch on deeper conceptual issues 
\cite{perc4},
the percolation picture of confinement misses the order of the 
phase transition both at $N_c=3$ and $N_c \rightarrow \infty$,
so its direct relevance to confinement is questionable. 

\begin{figure}[h]
\begin{center}
\includegraphics[width=18cm]{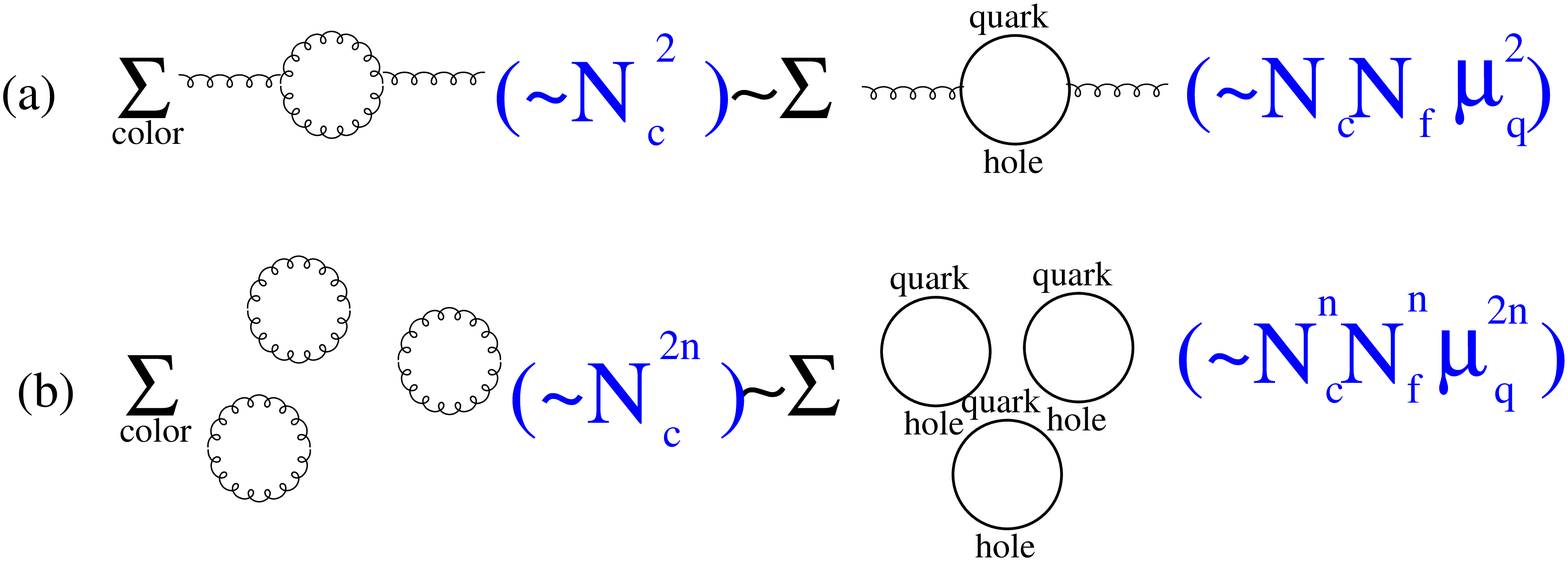}
\caption{ (color online)
  \label{diagrams}
 Panel (a) shows the interplay between anti-screening by gluons (driving confinement) and screening at high chemical potential.
  As the panel (b) shows, higher order corrections will not alter th dependence leading in $N_f/N_c$ \cite{fairness}.}
\end{center}
\end{figure}
There is however a newly conjectured regime where the percolation picture might be viable:
it is the proposed ``quarkyonic matter'' at low temperature (below the deconfinement temperature $T_c$)
and moderate density (one baryon per baryonic size, $\mu_Q = \mu_B/N_c \simeq \Lqcd$)
\cite{quarkyonic,kojo,quarkyonic2,quarkyonic3,quarkyonic4,quarkyonic5,quarkyonic6,quarkyonic7}, which is confined
(the excitations at the Fermi surface are baryonic) but ``quark-like'', in that pressure and perhaps also entropy density feel
the quarks below the Fermi surface and consequently scale as $N_c^1$, as opposed to $N_c^0$.
\begin{figure}[t]
\begin{center}
	\includegraphics[width=1.0\textwidth]{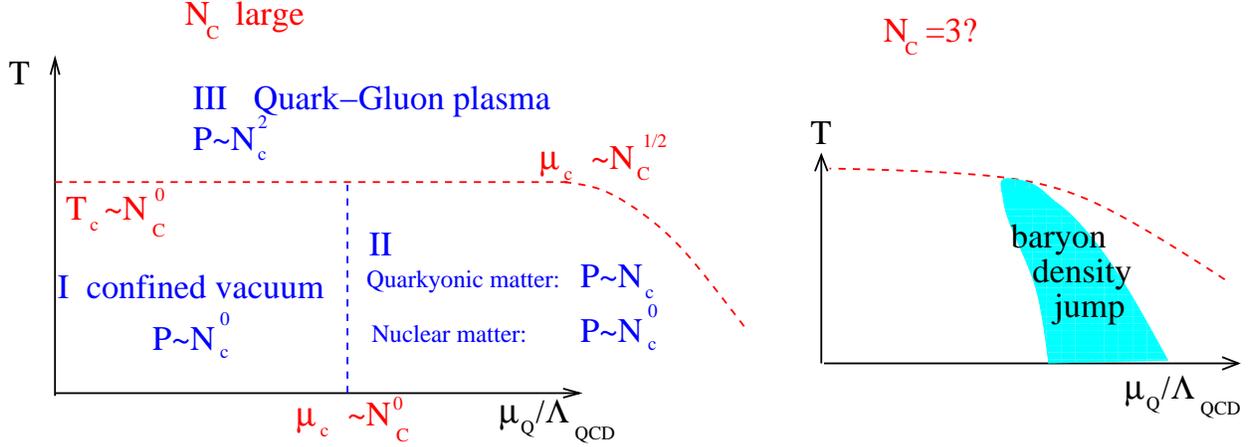}
	\caption{
		(color online) Larger left panel: The phase diagram in the $N_c \rightarrow \infty$ limit, where the deconfinement line becomes flat
		as quark corrections vanish, and a first order phase transition, with the baryon density as an order parameter,
		rises up vertically w.r.t. temperature at $\mu_Q \sim \lqcd$.
		The scaling of the pressure with $N_c$ is shown in the different phases.   Smaller right panel: a qualitative sketch of the expected situation at $N_c=3$.
		\label{figphase}
	}
\end{center}
\end{figure}

Unlike in \cite{perc1,perc2,perc3}, the ``quarkyonic'' transition is thought to be {\em distinct} from deconfinement,
to be found in lower energy heavy ion scans \cite{low1,low2,low3,low4} at low temperature but high baryo-chemical potential $\mu_B = N_c \mu_Q$
(a description of deconfinement at finite density as percolation was also postulated in \cite{greiner1,greiner2}).

The reason for conjecturing the existence of a new phase boils down to comparing the quark-hole screening with the gluon-gluon antiscreening at large chemical potential (Fig.~\ref{diagrams}):
confinement is broken when the screening by quark-hole pairs  $\sim \mu_Q^2 N_c N_f$ at the Fermi surface (which decreases the effective coupling)
overpowers anti-screening by gluon loops ($\sim N_c^2$), which drives the coupling constant above
non-perturbative values at momenta $\sim \Lqcd$ and ultimately causes the onset of the strong coupling regime.
This can be used to get an estimate for the low-temperature deconfinement point as scaling {\em at least} as $\sim \lqcd \sqrt{N_c/N_f}$.
A cursory examination of Fig.~\ref{diagrams} (bottom panel) shows that higher loops giving a $(N_c/N_f)^{z>1/2}$ scaling \cite{fairness}, and hence {\em perturbative} contributions to the QCD $\beta$-function, cannot lower the extra scale $\mu_q \sim \sqrt{N_c/N_f}^{z\geq 1/2} \lqcd$ which appears when one explores the deconfinement phase transition in chemical potential rather than temperature.

Thus, the phase diagram at $N_c \to \infty$ looks like the one in Fig.~\ref{figphase} (left) \cite{quarkyonic}:
the deconfinement line becomes infinitely flat.
At the same time, the transition to ``nuclear matter'', with the baryonic density as order parameter,
becomes infinitely sharp since the baryon mass $\sim N_c \lqcd$.
Therefore baryons drop out of the confined vacuum partition function entirely, but continue to be present at $\mu_Q \geq \lqcd$.

Hence, ``nuclear matter'' at $\mu_Q \sim \Lqcd$ should, at large $N_c$, be in the confined phase.
In configuration space, however, inter-quark distance $\sim N_c^{-1/3}$: for large enough $N_c$, then, one should be in the
confinement regime yet somehow neighboring quarks should be so close that asymptotic freedom applies.
The authors of \cite{quarkyonic} proposed a solution to this seeming contradiction by postulating matter in this regime is ``quarkyonic'',
with quark-like degrees of freedom deep inside the Fermi surface (and hence a scaling $\sim N_c^1$ for the pressure)
but baryonic excitations on the surface.

While the argument above is compelling, it raises somewhat subtle issues about how to characterize matter in the $\sqrt{N_c/N_f}\lqcd \geq \mu_Q \geq \lqcd$ part of the diagram.    Above $\mu_Q =\lqcd$, baryons will conceivably overlap.   If quarks are free within baryons, then how does one distinguish the ``quarkyonic phase'' from a deconfined phase?   Intuition from models such as the bag model \cite{bag} does indeed suggest that deconfinement happens at $\mu_Q \simeq \lqcd$, in contrast with Fig.~\ref{diagrams} and \cite{quarkyonic}.

It is clear that {\em if} color can flow within overlapping baryon regions and asymptotic freedom applies in the large $N_c$ limit, the Wilson loop expectation value within an area covering ``many overlapping baryons'' will break the area law due because, within the overlapping regions, the Gauge field configurations will fluctuate chaotically around a zero average\cite{wilson,feynman}.   
(an alternative way to see it, originally due to \cite{susskind}, is that if Gauge bosons can propagate through overlapping baryons, Gauss's law forces inter-quark fields to $\sim 1/r^2$).
In this scenario, quarkyonic matter will be essentially deconfined (according to the criteria set in \cite{wilson,feynman} to define confinement), and hence indistinguishable from a QGP, and the picture in \cite{bag} will be correct.
This is {\em possible}, since Fig.~\ref{diagrams} does not preclude {\em non-perturbative} contributions to the running of quark-quark interactions, which could in principle bring the critical $\mu_Q$ for deconfinement down to $N_c^0 \lqcd$.

However, as we explore in section \ref{sectheo} (and was explored in the past in the context of color superconductivity \cite{neut0,neut1,neut2,neut3}), it is not an inevitable conclusion: periodic quark wavefunctions, together with a generalization of spin-charge separation, can provide a physical mechanism whereby quantum numbers associated with quarks can move across arbitrary distances while color itself is confined to a configuration-space scale of $\sim 1$~fm.   
In this case, ``quarkyonic'' percolation and deconfinement are physically different phases, distinguishable by the usual order parameters associated with confinement.

Of course, it remains to be seen whether such mechanisms are realized in nature, and either possibility (a new phase or a non-perturbative breaking of $N_c$ scaling of the $\beta$-function) are interesting.
In the rest of this work we assume that dynamics of the type in section \ref{sectheo} holds, so a regime where color is localised, but quarks of neighboring baryons can interact is possible.    We then use the model developed in \cite{percolation1} to try to define where, in density, temperature and $N_c$, can this regime be located, in order to provide future experimentalists and phenomenologists tools to distinguish between the above possibilities.


The possibility of exploring the quarkyonic transition experimentally further assumes that physics at high chemical potential
is qualitatively the same when $N_c$ is varied from 3 to infinity.
In \cite{percolation1} it has been shown that for a wide variety of reasonable propagators at a fixed baryonic
number density of $\rho_B = \lqcde{3}/8$ a percolation transition is found as $N_c$ is varied.
If one identifies the percolation transition with the quarkyonic phase, deconfinement and percolation are indeed separate,
and they cover different regions not just in $T$ and $\mu_B$, but also along $N_c$.

In this work, we aim to extend the results of \cite{percolation1} to variable density and non-zero temperature.
The purpose of this exercise is to determine the role of percolation in the full $T$-$\mu$-$N_c$ phase diagram,
and to see whether percolation is involved in the physical $N_c=3$ world, or, instead,
whether this transition divides our world from the ``truly large-$N_c$'' regime.

Specifically we aim at determining whether there is a region, in the $T$-$\mu_B$ plane,
where a percolating yet confined phase is likely at $N_c=3$, \mbox{$\lqcd \leq \mu_Q \leq \sqrt{N_c/N_f} \lqcd$}.
If so,\footnote{\label{strfoot}In our world of course $N_f=2$ if the strange quark is heavy, and $N_f=3$ if the strange quark is light.
	Since the bare strange quark mass $\sim \lqcd$, it is far from clear which limit applies,
	yet this is the crucial question determining whether $N_f/N_c$ is an expansion parameter {\em at all}.  
	The mass of the strange quark might well be the crucial {\em qualitative} uncertain driving factor in our results,
	and hence fundamentally determining the nature of the QCD phase diagram \cite{janstrange,jansbook}.}
this would be the natural region to investigate for quarkyonic effects in experiment.
We also ask ourselves whether percolation is related to the more usual liquid-gas phase transition,
and if its onset therefore accounts for the large phenomenological failures of the large-$N_c$ picture in this regime \cite{vdw}.

We close with a discussion outlining what an effective theory for percolating matter would look like,
and suggesting ways of looking for it in both lower energy experiments and astrophysical searches (neutron and proto-neutron stars).

\section{The variable-density percolation model}
\label{secmodel}
The strategy used to investigate the percolation properties of high-density baryonic matter 
is a generalization of the model presented in \cite{percolation1}, to which we refer for further introductory details.
In our description, baryonic matter is arranged in a cubic lattice, with a baryon
sitting at each lattice site: its quarks will be randomly, independently positioned
according to a hard-sphere distribution with radius $1/\lqcd$:
$f(\mathbf{x}) \propto \Theta(1 - \lqcd|\mathbf{x}-\mathbf{x}^\mathrm{center}|)$.
In \cite{percolation1}, each sphere touches exactly its six neighbors, that is,
the lattice spacing is fixed to $2/\lqcd$ and the density $\overline{\rho}$ is therefore 
fixed (we will relate $\overline{\rho}$ to the thermodynamic baryonic density $\rho_B$ in Section \ref{secconf}).
Replacing the cubic arrangement with another regular 3D lattice would have changed the percolation threshold by
$\order{30\%}$ or so \cite{percthresholds}, and hence not impacted our results qualitatively.

The generalization to a variable-density setting
is realized by the introduction of the parameter $\epsilon$, defined as
the ratio of the lattice spacing over twice the spheres' radius: thus,
the density $\overline{\rho}_0$ examined in \cite{percolation1} had $\epsilon=1$.
Since at $\overline{\rho}_0$ each baryon occupies a volume of $(2/\lqcd)^3$, we now have
\eq
	\overline{\rho}(\epsilon) = \overline{\rho}_0\frac{1}{\epsilon^3} = \frac{\lqcd^3}{8}\frac{1}{\epsilon^3}\;\;.
	\label{eq:epsilon_rho}
\qe

In support of this classical, static description of baryonic matter, we keep an eye to the large-$N_c$ limit and note
that the propagation speed of Fermi-surface baryons, \\ $\sim1/\sqrt{N_c/N_f}$ in the confined regime, 
is parametrically smaller than the characteristic momentum of quarks, $\sim N_c^0$: hence, percolating quarks see the 
baryons as quasi-static (a crystal at larger $N_c$~\cite{witten,crystal}, and, presumably, a disordered ``glass'' at smaller $N_c$).
Deviations from the ``baryons are spheres'' assumption might become significant when the number of colors
approaches the number of neighbors of a densely packed system \cite{vdw}, which in 3D means $N_c \sim \order{10}$.
This, as we will show, coincides roughly with the percolation threshold for ``sensible'' choices of the parameters at play.

We note that we are assuming that baryon size does not depend on baryonic density.
Seemingly, this assumption is counter-intuitive since the pionic corona around the baryon is 
set by $f_\pi^{-1}$, which should decrease as chiral symmetry is partially restored \cite{brownrho}.
We note, however, that $f_\pi \sim \sqrt{N_c}$ and hence diverges for all chemical potentials as 
$N_c \rightarrow \infty$. In this limit baryons interact strongly with pions \cite{thooft,witten,heinzgiac},
as strongly in fact as with each other ($NN$ interactions via pions scale in the same
way as $NN$ interactions via quarks \cite{witten}).
The baryon size, however, stays finite and $\simeq \lqcd^{-1}$ in this limit.
This suggests that at large $N_c$ the baryon size becomes a lot more dependent on
``bag physics''/confinement (and in general the scale at which QCD becomes strong) than on the pion corona.
Because of this, pion size changes with $N_c$ are at best a subleading effect, to be disregarded here.
The fact that percolation, as defined here, is primarily an effect of confinement physics, 
as it survives in a world with only one quark flavor and no mesons, reinforces this conclusion.

Let us briefly summarize the results in \cite{percolation1} before generalizing to the present setting.
We assume a probability (``squared propagator'') for two quarks in different baryons to exchange energy/momentum
with the essential properties of (a) getting weaker as $\lambda/N_c$ at increasing $N_c$, and (b) dropping quickly to zero
around some confinement scale $r_T/\lqcd \propto \order{1} N_c^0$.
These requirements enforce the relevant physics of the problem; we then consider two representative choices for the ``propagator'',
inspired by the Gribov-Zwanziger theory \cite{conf2,conf3}: the step-function in coordinate space and the step-function
in momentum space (that is, the squared Fourier transform of a $p$-space theta-function).
These can arise out of chromo-field interactions such as in \cite{greiner1,greiner2}. 
Their expression, suitably normalized, is given by
\eqa
 F_T(y) &=&  \frac{\lambda}{N_c} \;
\Theta\Big(1-\frac{y}{r_T/\lqcd}\Big) \;\;; \label{eq:propagator_t} \\
	F_K(y) &=&  \frac{\lambda}{N_c} \; \frac{2r_T^2}{\pi y^2} \sin^2\Big( \frac{y}{r_T/\lqcd} \Big) 
		\quad
	\mbox{\footnotesize{
		$ \propto~~\left[  \int \de p \hspace{0.3em} e^{-ipy} \Theta(p-\lqcd/r_T) \right]^2 \;\;, $
	}}
	\label{eq:propagator_k}
\qea
respectively, where $y$ is the inter-quark distance in physical units.
The two parameters are $r_T \sim 1$ and $\lambda \sim 1$ ('t Hooft coupling), controlling respectively the range and the intensity
of the interquark exchanges.   

Conceptually, these definitions leave some ambiguity of what is ``propagated''.
Our working hypothesis is that Eqs.~\ref{eq:propagator_t} and \ref{eq:propagator_k} represent tunneling-driven interactions of quarks from different baryons
exchanging conserved quantum numbers (spin, flavor, energy-momentum) concurrently with some global color-neutralization mechanism acting at distance scales $\geq \lqcd^{-1}$.
We will leave discussion as to how this could happen to Section \ref{sectheo}.
We note, however, that similar propagators have already been used in the context of quarkyonic matter,
having been instrumental in the study of the conjectured quarkyonic ``chiral spirals'' \cite{spiral1,spiral2}
(the results in \cite{kojo} are based on a propagator of the form of Eq.~\ref{eq:propagator_k}).

With $F(y)$ and $f(\mathbf{x})$ as input, then, a probability $p$ for the exchange between neighboring baryons $A$ and $B$ is computed via
\eq
	p(N_c) = 1 - \Bigg[
		\int f_A(\mathbf{x}_A)\de^3\mathbf{x}_A
		\int f_B(\mathbf{x}_B)\de^3\mathbf{x}_B
		\Big( 1 - F(|\mathbf{x}_A-\mathbf{x}_B|) \Big)
	\Bigg]^{N_c^\alpha}\;\;;
	\label{eq:b-to-b-probability}
\qe
In \cite{percolation1} we concluded that, in order to meet the expectation $p \to 1$ for $N_c\to\infty$, the correct choice is
$\alpha=2$: we interpret it as a cross-baryon ``interaction'', as opposed to the cross-baryon ``propagation'' associated to $\alpha=1$.

Varying baryon density should not alter the $\alpha=2$ dependence assumed in \cite{percolation1}:
if $N_c$ is large enough, $\alpha=2$ will obviously always dominate over the $\alpha=1$ component.
Furthermore, as shown in \cite{percolation1}, $\alpha=1$  would imply a regime where  $N_c \rightarrow \infty$ matter would be less correlated (and hence less strongly bound) than $N_c \ll \infty$ (the nuclear binding energy dependence of $N_c$ would have a peak at some intermediate $N_c$ and reach the limit of \cite{witten} from above). Since quantum corrections generally make a many-Fermion system more repulsive (by Pauli blocking arguments alone), and since at large $N_c$ attractive channels dominate over repulsive ones,
  it is difficult to imagine a physical justification for such behavior,  neigher in naive $N_c$-counting nor Gauge/gravity.

At large $N_c$, therefore, it is natural to expect $\alpha=2$, but at $N_c=3$ an $\alpha=1$ component, negligible at $N_c \rightarrow \infty$, could be significant.
As the next section will make clear, assuming $\alpha=2$ throughout can be regarded as an ``optimistic limit''
for the existence of the percolating phase, and any $\alpha=1$ admixture will make the percolating phase less likely.

A similar discussion is needed to clarify the role of antibaryons in percolation at $T>0$.
Since by percolation we mean the delocalization of the quark wavefunction across baryons,
as we do in Section \ref{sectheo}, then only quarks delocalize, since delocalization is brought
about not by deconfinement but by the formation of a quark Fermi surface.
Antibaryons will then show up as an impurity in the percolation links.
We shall ignore this impurity for the current work, as it we are concentrating on the
``most optimistic scenario'' and any impurity will flatten the $\rho_B$-$N_c$ percolation curves,
making a percolating but deconfined phase less likely.
In addition, the  probability of having a local impurity $\sim \exp[-N_c] \sim 1\%$ even
at the highest temperature $T\simeq T_c$, so its effect should be smaller than other effects we neglect in this work.


The resulting $p$ is a function of $N_c$ (see \cite{percolation1} for details) and can be compared with the bond-percolation
threshold to determine if large-scale correlations occur or not, identifying a critical $N_c^*$ where the
system starts to percolate.
Note that direct exchange between non-neighboring baryons is neglected by construction.

However, in the variable-density setting, when $\epsilon$ is small we have substantial overlapping between the
spheres, and non-negligible contributions from non-nearest-neighbor direct exchange: we need to take into account
a set of baryon-to-baryon probabilities $p_i$,
one for nearest-neighbors, one for neighbors with relative distance of $(1,1,0)$ spacings, and so on.
In practice, we considered nine probabilities $\{p_i, i=0,\cdots,8\}$, associated to nine ``neighboring classes''
(we assume the probability of direct exchange between spheres that are further apart can safely be neglected):
the corresponding relative distances, in units of the interbaryonic distance, read:
\eq
	(1,0,0),(2,0,0),(1,1,0),(2,1,0),(2,2,0),(1,1,1),(2,1,1),(2,2,1),(2,2,2)\;\;.
\qe
\begin{figure}[h]
\begin{center}
	\includegraphics[width=17cm]{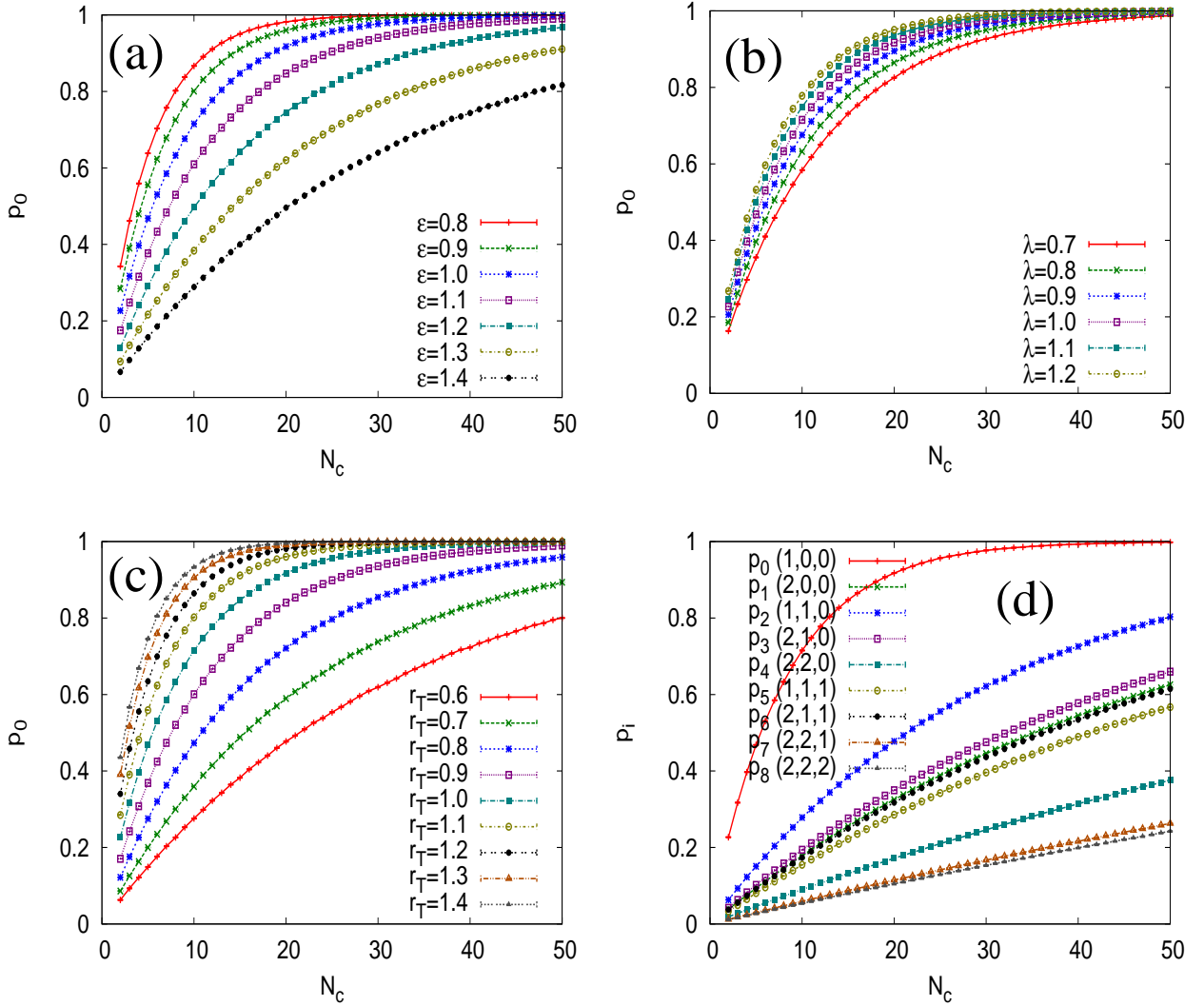}
	\caption{(color online) 
		Baryon-to-baryon probabilities for the $F_K$ propagator;
		the reference case is $\epsilon=r_T=\lambda=1.00$.
		(a): $p_0$ for varying $\epsilon$.
		(b): $p_0$ for varying $\lambda$.
		(c): $p_0$ for varying $r_T$.
		(d): $p_i$ at all neighboring-classes considered.
	}
	\label{fig:samplecurves}
\end{center}
\end{figure}

Thus, for each choice of the parameters $(r_T, \lambda)$, each ``propagator'' and each $\epsilon$,
a set of $\{p_i\}$ can be evaluated numerically with Eq.~\ref{eq:b-to-b-probability}, by
placing the two centers at the desired distance.
Representative results are shown in Fig.~\ref{fig:samplecurves}
for the $F_K$ propagator.
We examine 9 values of $r_T \in [0.6$:$1.4]$, 6 of $\lambda\in[0.7$:$1.2]$, and 13 of $\epsilon\in[0.8$:$1.4]$:
the latter corresponds to ranging in density from $\sim0.046\,\lqcde3$ to $0.244\,\lqcde3$.
We generate data for all integer $N_c$ from 2 to 80.  The $\{p_i\}$ are
calculated numerically by sampling the integrand about $4\cdot10^5$ times per setup.

The next step is to translate the set $\{p_i\}$ to a single probability $p$, which will be compared to the
3D cubic-lattice bond-percolation threshold $p_c \simeq 0.2488$.
This is accomplished by means of Monte Carlo renormalization (see, e.g., Chapter 4 of \cite{stauffer}):
we perform a blocking step on a cell of $b^3$ sites, mapping the problem to a super-lattice whose sites
are the corners of the cell; the corresponding $p$ is evaluated by numerical simulation, with the $\{p_i\}$
as input, as the probability that two opposite planes on the cell are connected by a continuous path.
Since $p_0$ is the only relevant coupling in the blocking-out flow, we are effectively moving
on a RG trajectory in the nine-dimensional space of the $p_i$, whose fixed point -- the percolation critical point --
lies on the axis $p_i=0,i>0$.
\begin{figure}
\includegraphics[width=8.cm]{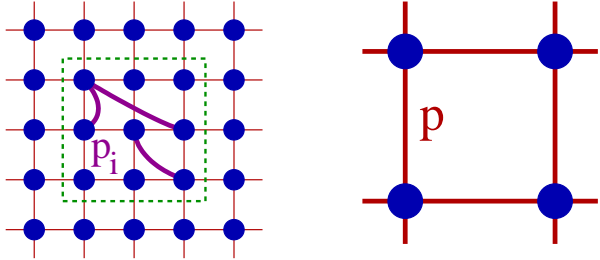}
        \caption{(color online) 
			Schematic procedure for the RG-step. The $b=3$ cell on the left, with all its bond probabilities $p_i$,
			becomes an elementary square of the super-lattice on the right,
			with the associated $p$ as its only, nearest-neighbor, bond probability.
		}
		\label{fig:blocking_out}
\end{figure}
The procedure is illustrated in Fig.~\ref{fig:blocking_out}.
In practice, we start with an empty $b^3$ lattice and, considering all pairs of sites,
we switch on the links according to the $p_i$ for the corresponding neighboring class.
At the end of the process we can have a continuous path connecting the $z=0$ wall of the cube
with the $z=b-1$ wall: the probability $p(p_i)$ for this to happen is computed by
repeating the operation many times (in our case, 20 thousand times per setup).

The exact choice of the ``crossing rule'' $R_1$ (that is, we get a 1 if and only if there is a connection between 
two opposite wall in a chosen direction, ignoring what happens along the other two directions) is arbitrary:
for $b$ large enough, all recipes would lead to the same final result.

In the limit $b\to\infty$, the crossing probability is a step function, zero before percolation and one after percolation. When 
evaluated exactly at criticality, it assumes a universal value depending only on the crossing rule considered. In three dimensions,
with the $R_1$ rule, $p^*(b) = \Pi_1 + \beta b^{-y}$
 with the limit value  $\Pi_1=0.265$ \cite{nicolai}, and leading correction $b^{1/\nu}$, with $\nu\simeq 0.8765$ the 3D percolation critical exponent.

\begin{figure}
\begin{center}
	\includegraphics[width=9.8cm]{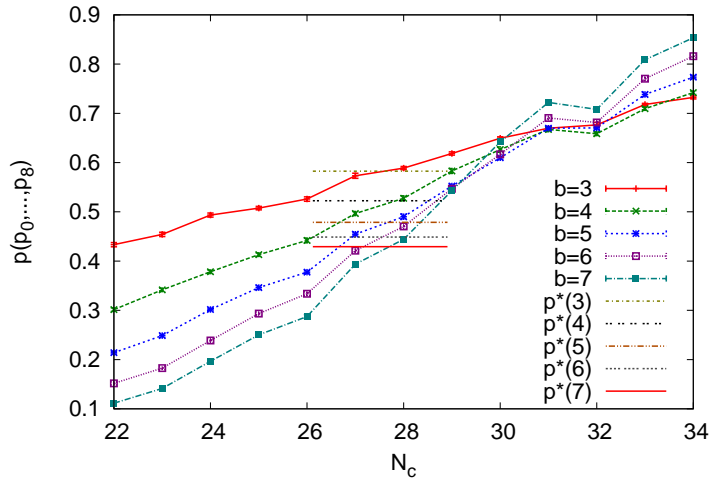}
	\caption{(color online)
		Determination of the $b$-dependent critical $N_c^*$, for $F_T$ with $\epsilon=0.8$, $\lambda=0.7$, $r_T=0.6$. Horizontal lines mark $p^*$.
	}
	\label{fig:multip}
\end{center}
\end{figure}

In order to check the implied assumption of a $b$ ``large enough'',
cell sides from 3 to 7 lattice steps are employed, and the results compared for stability.
This technique is first tested on the one-dimensional subspace of nearest-neighbor-only $p_0 \to p$,
where we roughly reproduce the scaling to the large-$b$ limit: yet, the curves $p_b(p_0)$
have a rather $b$-dependent value $p^*$ at the critical threshold:
\eq
		p^* = \{0.5825,0.5226,0.4788,0.4488,0.4293\}\;,\mbox{for}\;b=3,4,5,6,7\;\;.
\qe
This leads us to define an effective, $b$-dependent critical $N_c^*$,
the value at which the full renormalization step yields the $p^*$ for that cell size (practically found by interpolation).
We are satisfied with the procedure whenever $N_c^*$ shows a substantial independence of $b$;
indeed, this will be the case: for example, at $\epsilon=0.8$, $\lambda=0.7$ and $r_T=0.6$,
we get for the propagator $F_T$ (cf.~Fig.~\ref{fig:multip}):
\eq
	N_c^* = \{27.9, 27.6, 27.9, 27.6, 27.5\}\;,\;\;b=3,4,5,6,7\;\;.
\qe
The final result is then obtained by the above population of results.
The choice of considering bonds between baryons far apart has, however, a drawback: namely, the propagator $F_K$ (Eq.~\ref{eq:propagator_k}),
as it is, has local maxima around $y\lqcd \sim \pi r_T, 3\pi r_T$ and so 
on, which dramatically alters the results in a nonphysical way; the 
coincidence of the secondary peaks with neighboring positions is of course 
specific to a cubic lattice, and should not occur in a disordered 
''glass of baryons''.

More generally, however, the power law ``tail'' in the propagator makes 
correlation lengths diverge at $N_c \sim \order{1}$ for 
densities arbitrarily close to $\order{\lqcd^3}$).
The implementation of such a sharp step propagator in momentum space, however, 
contradicts our intuition of statistical physics:
a probe charge interacting with a statistical medium is generally exponentially screened by charge-hole pairs,
acquiring a screening length of dimension $\sim 1/\mu$ (or $\sim 1/T$ for $T \sim \mu$).
This is a ``universal'' feature of interacting systems at equilibrium, since any backreaction of a charge on a field,
to first order, will give a similar effect (the absence of such a screening leads to a divergence in the partition function of the fully interacting system, the hydrogenic atom's case is a well-studied example of these classes of phenomena \cite{hydpart1,hydpart2}). 
Since $\mu_Q \sim \Lqcd$ in the limit we are using,
the propagator should be exponentially screened at a similar scale.
Then, in the following, we will use a ``K''-propagator altered by an exponential screening:
\eq
	F_S(y) = \frac{\lambda}{N_c} \frac{2r_T^2}{\pi y^2} \sin^2\left( \frac{y \Lqcd}{r_T} \right) \cdot e^{-M |y|}\;\;,
	\label{eq:propagator_s}
\qe
that is, a $F_K$ times a damping exponential factor whose characteristic length is, in practice, fixed to $1/\Lqcd$.
$F_S$, taking the square root and antitransforming, yields
\begin{displaymath}
	\widetilde{g_S}(k)=
		\int_{-\infty}^{+\infty} \de x \sqrt{F_s(x)} e^{ikx}
				\propto
					\left\{ \mathrm{atan}\left[\frac{2\lqcd}{M
							r_T}\left(k\frac{r_T}{\lqcd}+1\right)\right]
					- \mathrm{atan}\left[\frac{2\lqcd}{M
							r_T}\left(k\frac{r_T}{\lqcd}-1\right)\right]
					\right\}\;\;,
\end{displaymath}
i.e.~a ``rounded off'' step function in momentum space.

Following the above methodology, we obtain curves in $\epsilon$, one for each propagator ($F_T$, $F_K$ -- later ignored -- and $F_S$) and 
each $(r_T,\lambda)$, that can be translated into curves $N_c^*(\overline{\rho})$ through Eq.~\ref{eq:epsilon_rho}. 

\begin{figure}
\begin{center}
	\includegraphics[width=1.\textwidth]{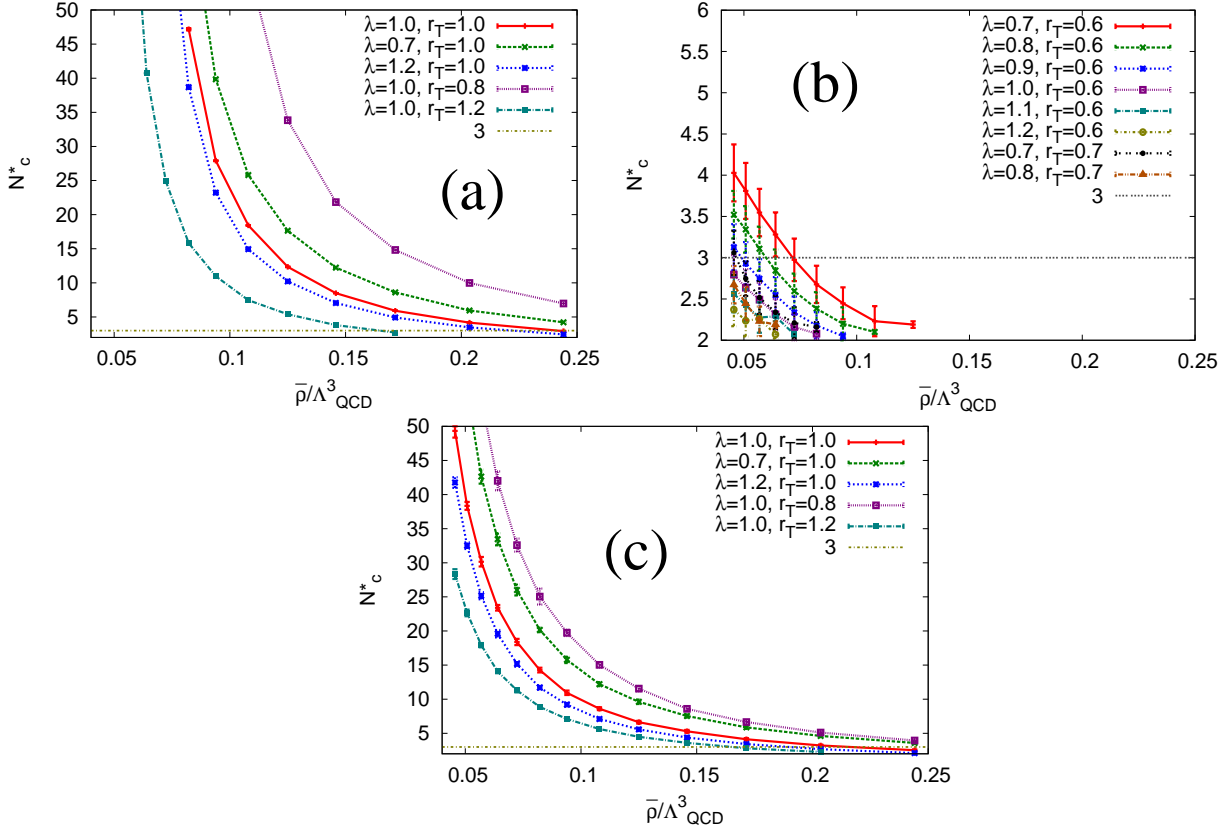}
	\caption{(color online) 
		Representative results, with some choices of $r_T, \lambda$, for the critical percolation $N_c^*$
		as a function of the baryon density, coming from using the propagators $F_T$ of
		Eq.~\ref{eq:propagator_t} (a), $F_K$ of Eq.~\ref{eq:propagator_k} (b) and $F_S$ of Eq.~\ref{eq:propagator_s}
		(c).
	}
	\label{fig:nc-rho-perc}
\end{center}
\end{figure}

Examples of such curves are shown in Fig.~\ref{fig:nc-rho-perc}.
We note they are well fitted to the form
\eq
	N_c^*(\overline{\rho}) = \frac{A}{(\overline{\rho}^Z+B)}\;\;.
	\label{eq:perc_parametrisation}
\qe
We have shown that the percolation transition line in $\overline{\rho}$-$N_c$ space is strongly curved.
This has the potential of dramatically altering the conclusions of \cite{percolation1},
provided the assumption that the relevant density is $\sim
\lqcde{3}/8$ is relaxed:  at a greater density, provided baryons still
exist,
$N_c^*$ can very well be lowered from $\order{10}$ to $N_c=3$, making the percolating phase accessible to experiment.
The existence of baryonic states, of course, implies that the relevant density is still in the confined phase.
In Section \ref{secconf}, therefore, we examine the interplay between the confining and percolating transitions.

\section{The confinement phase transition in $N_c$-$\rho_B$-$e$ space}
\label{secconf}
As discussed in the introduction, if one identifies the density of interest for percolation, $\rho_B \sim \lqcde{3}/8$
(one baryon per baryon size), with the thermodynamic density, then, at $T \ll T_c$, the percolation transition happens firmly in the confined phase in the large-$N_c$ limit because of the $\mu_Q^2$ dependence of the Quark-hole screening diagram (Fig. \ref{diagrams}).

We also know that, at $T \simeq T_c$ -- where the chemical potential for deconfinement is zero -- does not scale with $N_f,N_c$.   Hence,
\eq
	\left. \mu_Q^{\mathrm{conf}} \right|_{T \ll T_c} =\order{1} \sqrt{\frac{N_c}{N_f}} \Lqcd \;;\;\;
	\left. \mu_Q^{\mathrm{conf}} \right|_{T \sim T_c} \rightarrow 0 \;;\;\;
	\left.T_c\right. \simeq \left(\frac{2\lqcd}{3}\right)N_c^0 N_f^0
		\label{munclims}
\qe
In this Section, we will use the ideal gas Ansatz for the nuclear liquid to translate these estimates
into an estimate of how $\rho_B^\mathrm{conf}$ and the energy density $e^\mathrm{conf}$ depend on $N_c$.
We note that, even in the case of Van der Waals corrections of order $\order{N_c}$, the density $\rho_B$ at a given $T$
will change at most with terms of $\order{1}$ \cite{vdw}, therefore can be ignored for this calculation.

We start with the ideal gas formula for the density of a relativistic massive 
gas of fermions\footnote{For ``baryons'' to be fermions, we should limit ourselves to integer, odd $N_c$:
as long as we don't include the excited spin states, however, the formulas are valid - and have
here been calculated - for generic positive values of $N_c$.}
\eq
	\rho_B =\statfactor \left( B_1 (m,T,\mu_B) -B_1 (m,T,-\mu_B) \right)\;\;,
	\label{eq:density_first-equation}
\qe
the first term accounting for baryons and the second for antibaryons, with \cite{thesis,jansbook}
\eqa
	B_1(m,T,\mu_B) &=&  \int_0^\infty \frac{p^2 dp}{\exp \left[ \frac{1}{T}\left( \sqrt{p^2+m^2}-\mu_B \right) \right] +1} = \\
		&=& \sum_{n=1}^{\infty} (-1)^n m^2 \frac{T}{n} \exp \left( \frac{n \mu_B}{T} \right) K_2 \left(\frac{n m}{T} \right) \nonumber
\qea
and $K_n(x) = \int_0^\infty e^{-x \cosh t} \cosh(n t) \de t$ the modified Bessel function.   
The degeneracy terms $g_f$ and $g_s$ are both somewhat non-trivial:
$g_f$ counts the total ``generalized isospin'' states accessible allowed for the baryon, and so is $1$ for one flavor, $N_f(N_f-1)$ for a higher number.
If we ignore excited spin states, the baryon will always have spin $1/2$, so that $g_s(N_c)\equiv2$;
we will later refine this simplification.

For $T \ll N_c \lqcd$ the antibaryon term becomes negligible and the baryon density becomes, in units of ``one baryon per baryon size'':
\eq
  \rho_B = K_0^{3/2} \frac{4}{3} \pi (~0.6\mathrm{~fm}~)^{-3}  \eqcomma \mu_B \simeq 2\pi \left( \frac{3\sqrt{\pi}}{4} \right)^{2/3} \frac{\rho_B^{2/3}}{ m}
  \eqcomma  \mu_B \simeq 1.7 K_0 \mathrm{~GeV}\;.
\qe
For $K_0=1$ (approximately one baryon per baryon size), 
this corresponds to the density examined in \cite{percolation1},
where $N_c \sim \order{10}$ is necessary for percolation.
We need therefore $K_0 >1$, but not enough to trigger deconfinement.

We note that it is impossible to estimate the $\order{1}$ parameters in Eq.~\ref{munclims},
and hence $K_0$, to better than an order of magnitude: these factors depend on the $SU(N_c)$
structure constants showing up in the two diagrams of the top panel of Fig.~\ref{diagrams},
as well as the mean field corrections to the quark and gluon wavefunctions, which enter in the
incoming and outgoing lines in the same top panel of Fig.~\ref{diagrams} \cite{vogel}.
The latter are completely undetermined, even for a dilute gas at large $N_c$ \cite{cohenbar1,cohenbar2}.
Finite temperature and antibaryons will introduce additional modifications of $K_0$.

Unfortunately, this uncertainity radically limits the predictive power of this section,
since, as we shall see, factors of $\order{1}$ are crucial for deciding whether
a percolating phase does in fact occur at $N_c=3$.
Nevertheless, we shall continue to illustrate the issues at hand.

\begin{figure}
\begin{center}
	\includegraphics[width=13cm]{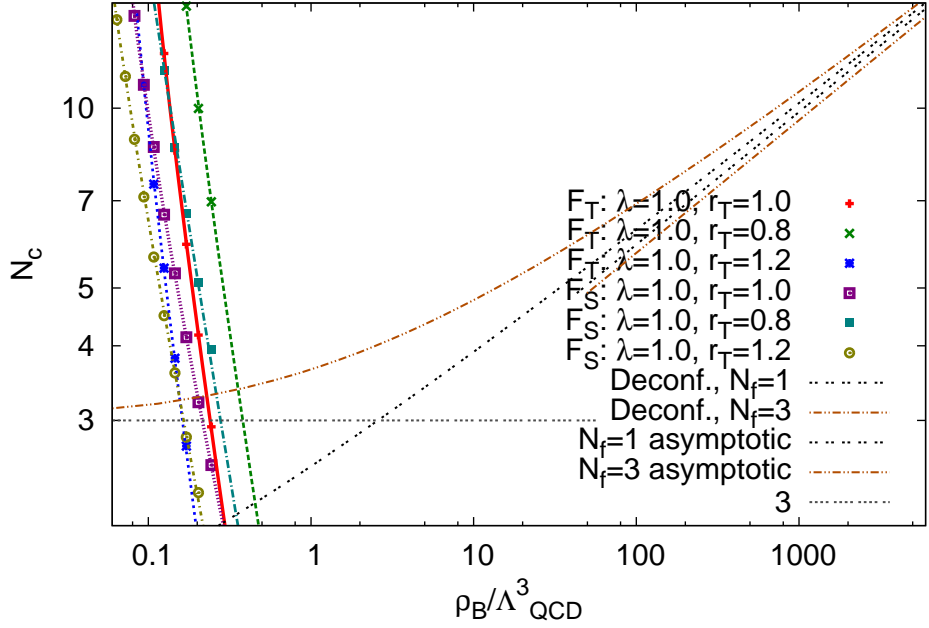}
	\caption{(color online) 
		Log-log plot of zero-temperature deconfinement curves versus percolation curves in the plane of $\rho_B \simeq \overline{\rho}$ and $N_c$.
		The deconfinement curves (Eq.~\ref{ncritdec}) are calculated for $N_f=1,3$ and are compared to their large-$N_c$ asymptotic form.
		The percolation curves are shown with their parametrization from Eq.~\ref{eq:perc_parametrisation}.
	}
	\label{percdec0}
\end{center}
\end{figure}

From Eq.~\ref{munclims}, we parametrize the zero-temperature baryonic chemical potential needed for deconfinement as
\eq
	\mu_0 = \mu^\mathrm{conf}_B(T \ll m) = \order{1} N_c^{3/2}N_f^{-1/2}\lqcd
\qe
(roughly $N_c^{1/2} N_f^{-1/2}$ baryons need to overlap), and the baryon mass is $m \sim N_c \Lqcd$.  
Omitting such factors of $\order{1}$, we are led to change the variables to
\eq
	\gamma = \frac{\sqrt{N_c}}{\mu_0}m  \;\;;\;\;
	\alpha = \frac{\sqrt{N_c}}{\mu_0}p  \;\;;\;\;
	\beta = \frac{\lqcd}{T}  \frac{N_c}{\sqrt{N_f}}\;\;;
\qe
note that at zero-temperature deconfinement we have $\beta \to \infty$ and $\gamma = \sqrt{N_f}$.
We write the critical $\rho_B^\mathrm{conf}$ for confinement as:
\eq
	\label{ncritt0}
	\rho_B^{\mathrm{conf}} \left( \beta\gg \lqcde{-1} \right)= \statfactor \frac{N_c^{3}}{N_f^{3/2}}  \lqcde{3}  B_2(N_c,\gamma,\beta)\;\;,
\qe
where
\eqa
	B_2 (N_c,\gamma,\beta) &=& \int \frac{\alpha^2\de \alpha}{1+\exp\Big[\beta\Big(\sqrt{\alpha^2+\gamma^2}-\sqrt{N_c}\frac{\mu}{\mu_0}\Big)\Big]}
		-\int\{\mbox{same with $\mu\to -\mu$}\} = \nonumber \\
	&=& \sum_{n=1}^\infty (-1)^n \frac{ n\gamma^2}{ \beta}  \sinh\left( \left( \sqrt{N_c}\frac{\mu}{\mu_0} \beta \right)^n \right) K_2 \left( n \gamma \beta \right)\;\;.
\qea
This is the general expression for the baryon density at deconfinement as a function of $N_c$.
For very low temperatures ($\beta\to\infty$, $\mu=\mu_0$), the second term in $B_2$ vanishes and the exponential at the bottom of the first term
is infinity if $\chi >0$ and zero if $\chi <0$, where
\eq
	\chi = \sqrt{\gamma^2 + \alpha^2} - \sqrt{N_c} = \sqrt{\alpha^2 + N_f} - \sqrt{N_c}\;\;.
\qe
We then get easily the low-temperature behavior:
\eq
	\rho_B^{\mathrm{conf}} (T=0)=  \lqcde{3}
	\frac{1}{6\pi^2} g_f g_s \frac{N_c^3}{N_f^{3/2}} (N_c-N_f)^{3/2} \sim \frac{g_f g_s}{6\pi^2} \lqcde{3} \frac{N_c^{9/2}}{N_f^{3/2}}\;\;.
	\label{ncritdec}
\qe
At $T=0$, ground-state baryons are the {\em only} possible hadronic degrees of freedom of the system.
Hence, one can trivially identify $\rho_B$ with the $\overline{\rho}$ of Section \ref{secmodel}, and directly compare deconfinement with percolation:
this is done in Fig.~\ref{percdec0}.
In contrast to what we find in \cite{percolation1}, it seems that a confined but percolating density at $N_c=3$ is possible.
The discussion in Section \ref{secmodel} elucidates what \cite{percolation1} missed: because of the curvature
of the density in $\rho_B$-$N_c$ space, the critical $N_c$ drops very rapidly with density, while the density required for deconfinement rises with $N_c$.
However, the conclusion made in \cite{percolation1} and \cite{vdw} still stands in that the densities required for it are well away from normal nuclear density,
as in the large-$N_c$ limit.
Hence, percolation is well distinct from the nuclear liquid-gas phase transition and might not arise if the strange quark is ``too light'' (see the footnote \ref{strfoot}).
In order to extend our knowledge to nonzero temperature, we choose the simplest parametrization for the deconfinement line in the $T$-$\mu_B$ plane, i.e.~that
of a quarter of ellipse whose radii are given by the known points $T=T_c$ and $T=0$:
\eq
	1- \theta^2 = \left(\frac{\mu^{\mathrm{conf}}_B}{\mu_0}\right)^2 \eqcomma \theta = \frac{T}{T_c} \simeq \frac{3}{2}\frac{T}{\lqcd}
	\label{ellipse}
\qe
Considering that the transition line is given by the interplay of the matrix elements shown in Fig.~\ref{diagrams}
this elliptical parametrization is actually physically well-motivated,
although it misses the $\sim T \mu$ interference between screening and anti-screening.

\begin{figure}
\begin{center}
	\includegraphics[width=17.cm]{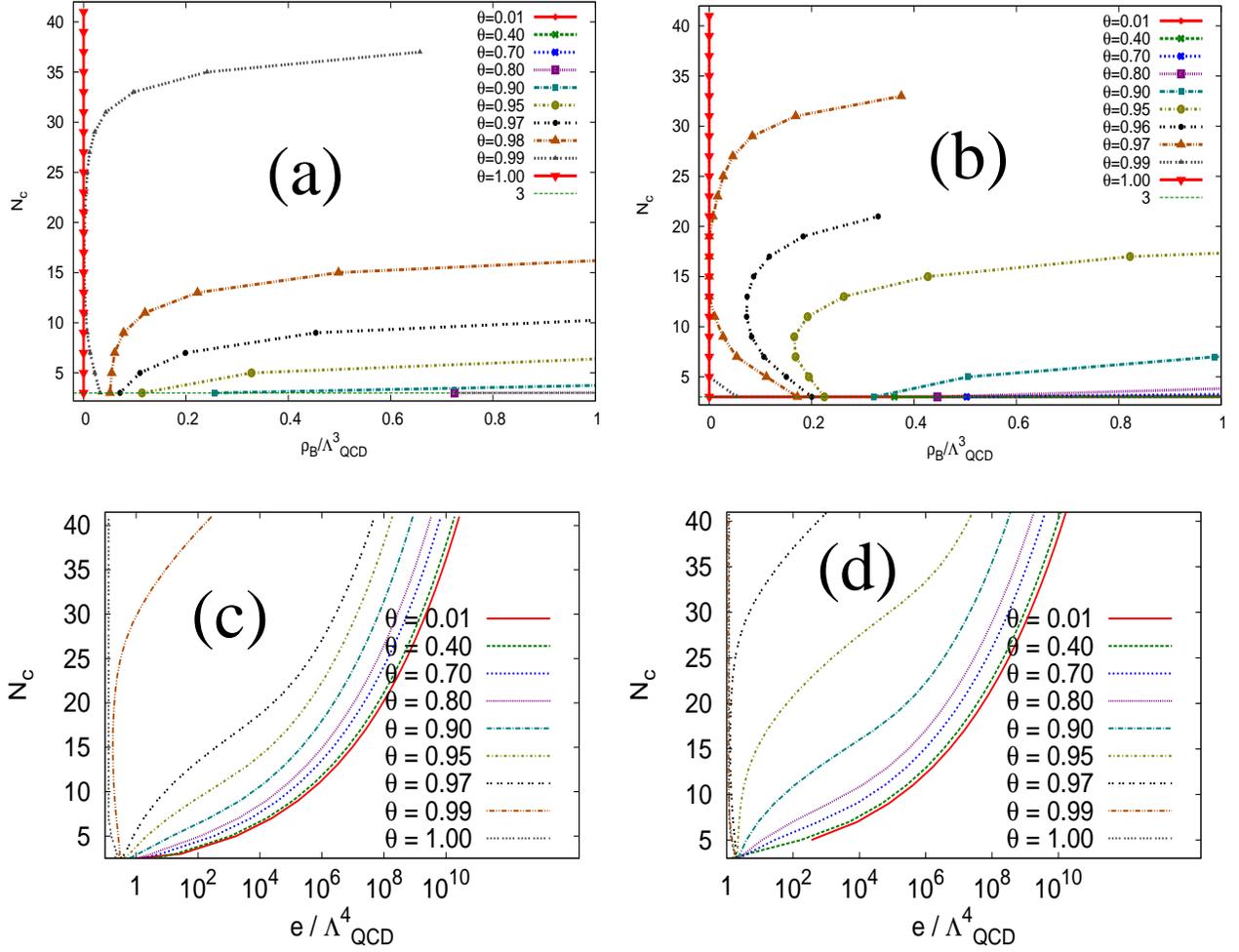}
	\caption{(color online)
		(a,b): deconfinement line in the $\rho_B$-$N_c$ plane according to Eq.~\ref{rhobT}.
               (c,d): deconfinement line in the $e$-$N_c$ plane from
Eqs.~\ref{rhoeconf}, \ref{eq:endensity}.		
		The plots are for $N_f=1$
			(a,c) and $N_f=3$ (b,d), and for various temperatures spanning $0<T\leq T_c$ (denoted by their $\theta$).
	}
	\label{percdec}
\end{center}
\end{figure}

When raising the temperature, we should also include states of higher spin in our calculation:
in the large-$N_c$ limit, a spin flip has a cost of $\sim \lqcd/N_c$ \cite{manohar}, and there
can be up to $(N_c-1)/2$ of them.\footnote{Admittedly, here we prefer to keep the model simple despite the fact that in our $N_c=3$ world
the cost of flipping, say, a proton into a $\Delta^+$ is around $\Lqcd$ and not $\Lqcd/3$.}
We now assume $N_c=2Q+1$ odd integer, and introduce a sum over spin states parametrized by $\eta=0,1,\cdots,Q$
(in this way, neglecting higher-spin states amounts to limiting all sums to $\eta=0$),
each carrying its degeneracy $(2\eta+2)$: this setup replaces the factor $g_s$ of Eq.~\ref{eq:density_first-equation}.
Thus we write, for nonzero temperature and including higher spins, the density as
\eqa
	\rho_B^{\mathrm{conf}} &=& \statfactornogs \frac{N_c^3}{N_f^{3/2}} \lqcde{3} \sum_{\eta=0,1\ldots}^Q (2\eta+2) \Bigg\{  	\label{rhobT} \\
		& & \qquad\phantom{-} \int \frac{\alpha^2\de\alpha}{1+\exp\Big[ \frac{3}{2}\frac{N_c}{\sqrt{N_f}}\frac{1}{\theta}
			\Big( \sqrt{\alpha^2+N_f} + \eta\frac{\sqrt{N_f}}{N_c^2} - \sqrt{N_c}\sqrt{1-\theta^2} \Big) \Big]} \nonumber \\
	& & \qquad - \int \frac{\alpha^2\de\alpha}{1+\exp\Big[ \frac{3}{2}\frac{N_c}{\sqrt{N_f}}\frac{1}{\theta}
			\Big( \sqrt{\alpha^2+N_f} + \eta\frac{\sqrt{N_f}}{N_c^2} + \sqrt{N_c}\sqrt{1-\theta^2} \Big) \Big]} \nonumber
	\Bigg\}\;\;.
\qea
This relation is plotted in Fig.~\ref{percdec}, top panels.
As temperature and $N_c$ rise, less and less energy density is carried by baryons, since the hadronic degrees of freedom are
light mesons (of mass $\leq 2 \lqcd$) carrying no baryonic quantum number, and heavy baryons ($\sim N_c \lqcd$).
The critical energy density $e^{\mathrm{conf}}$, neglecting the meson mass, is
\eq
	e^{\mathrm{conf}} =  N_f^2 \frac{ \pi^2}{15} T^4 + e_B^{\mathrm{conf}}\;\;,
	\label{rhoeconf}
\qe
with the baryonic contribution given by:
\eqa
	e_B^{\mathrm{conf}} &=& \statfactornogs \frac{N_c^4}{N_f^{2}} \lqcde{4} \sum_{\eta} (2\eta+2) \Bigg\{ \label{eq:endensity} \\
	& & \qquad\phantom{+} \int \frac{\alpha^2\Big[\sqrt{\alpha^2+N_f}+\eta\frac{\sqrt{N_f}}{N_c^2}\Big]\de\alpha}
		{1+\exp\Big[ \frac{3}{2}\frac{N_c}{\sqrt{N_f}}
		\frac{1}{\theta}\Big( \sqrt{\alpha^2+N_f} + \eta\frac{\sqrt{N_f}}{N_c^2} -
		\sqrt{N_c}\sqrt{1-\theta^2} \Big) \Big]} \nonumber \\
	& & \qquad + \int \frac{\alpha^2 \Big[\sqrt{\alpha^2+N_f}+\eta\frac{\sqrt{N_f}}{N_c^2}\Big]\de\alpha}
		{1+\exp\Big[ \frac{3}{2}\frac{N_c}{\sqrt{N_f}}
		\frac{1}{\theta}\Big( \sqrt{\alpha^2+N_f} + \eta\frac{\sqrt{N_f}}{N_c^2} + \sqrt{N_c}\sqrt{1-\theta^2}
		\Big) \Big]} 
		\Bigg\} \nonumber
\qea
In the $e$-$N_c$ plane, at $T_c$, the deconfinement line
\eq
e^\mathrm{conf}(T_c) = N_f^2 \frac{ \pi^2}{15} T_c^4
\qe
(with $T_c = 165 $ MeV or so \cite{latt1,latt2}),
is independent of $N_c$ (a vertical line), with a mixed phase in the region
\eq
 N_f^2 \frac{ \pi^2}{15} T_c^4  < e <   \left( N_c^2+ \frac{7}{8}N_c
N_f \right) \frac{ \pi^2}{15} T_c^2\;\;.
\qe
At $T\to 0$ (that is, $\theta\to 0$), the line is found by solving Eq.~\ref{eq:endensity}, and similar to the
plane shown in Fig.~\ref{percdec0} -- with all energy carried by baryons -- in this case we have $e=e_B$.
For intermediate temperatures, solving Eq.~\ref{eq:endensity} into Eq.~\ref{rhoeconf} will give an intermediate solution,
with a non-trivial approach to the $T=T_c$ case for different $N_c$: see Fig.~\ref{percdec}, bottom panels.
The behavior of the energy density is the reason why all curves ``curve'' (anticorrelate)
for low-to-moderate $N_c$ values in $N_c$-$\rho_B$ space: at moderate $N_c$ (including our $N_c=3$ world) baryons carry a
non-negligible fraction of the energy density even in the vacuum phase, with the flavor and spin degeneracy factors beating thermal suppression.
At high $N_c$, while of course baryons continue to carry the baryonic number, mesons carry the bulk of the energy density.
Since the scaling with $N_c$ of the deconfinement line is very
sensitive to where one is in $e$-$\rho_B$
(as per Eq.~\ref{ellipse}), this interplay can, at lower $N_c$, change
the $N_c$-$\rho_B$ correlation of confinement into an
anticorrelation.

At $T>0$ baryonic density and energy density are not the parameters driving the percolation phase transition anymore.
The way percolation treats baryons in Section \ref{secmodel} does not distinguish between baryons created through chemical potential,
whose number is conserved on average, and baryons created in pairs, whose number fluctuates.
Each baryon can give rise to quark tunneling and therefore participate in percolating links.

Therefore, and consistently with the discussion in \ref{secmodel}, recast the deconfinement curve in the $\overline{\rho}$-$N_c$ plane, where from Eq.~\ref{rhobT} we define:
\eq
	\overline{\rho} = \frac{g_f}{2\pi^2} \frac{N_c^3}{N_f^{3/2}} \lqcde{3} \sum_{\eta}
		\int \frac{(2\eta+2)\alpha^2\de\alpha}{1+\exp\Big[ \frac{3}{2}\frac{N_c}{\sqrt{N_f}}\frac{1}{\theta}
		\Big( \sqrt{\alpha^2+N_f} + \eta\frac{\sqrt{N_f}}{N_c^2} - \sqrt{N_c}\sqrt{1-\theta^2} \Big)}\;\;.
	\label{rhop}
\qe 
At $T=0$, where the antibaryon density is strictly zero, these distinctions are insignificant, and the comparison in Fig.~\ref{percdec0} suffers no problem.
At $T>0$, however, in principle we need to use the temperature and chemical potential to calculate the relevant non-conserved quantity.
\begin{figure}
  \includegraphics[width=1.1\textwidth]{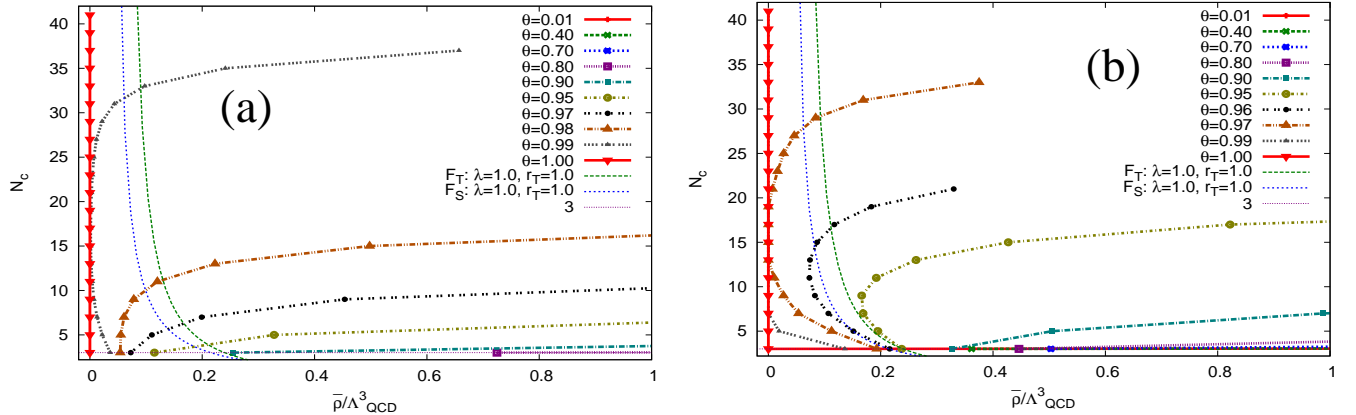}
       \caption{(color online) Percolation and deconfinement as a function of $\overline{\rho}$
				(defined in Eq.~\ref{rhop}), for one flavor (panel (a)) and three flavors (panel (b))}
       \label{percdecrhop}
\end{figure}
The final result is shown in Fig.~\ref{percdecrhop}.
As can be seen, the deconfinement line on the new axis is
quantitatively very similar to the upper two panels in Fig. \ref{percdec}.
Then, the two cases in the previous paragraph yield virtually
identical regimes on the phase diagram.

From Figs.~\ref{percdec0} and \ref{percdecrhop} we see that, in the $T$-$\mu_B$-$N_c$ space, there can be three distinct phases:
confined, deconfined, and confined but percolating.
In the latter case, arising at high $N_c$, the Polyakov loop expectation value is still zero and baryons are still physical degrees of freedom but
$\ave{q(x) \overline{q}(x')}$ should not vanish at scales larger than the baryon size due
to tunneling-driven quark interactions across asymptotically large distances.
At low $N_c$, the confinement density is lower than the percolation density.
Since the percolation transition necessitates baryons as physical states,
it therefore does not occur, and quarkyonic phases such as in \cite{quarkyonic} are not realized.

The critical $N_c$ allowing for a distinction between deconfinement and percolation is typically $\order{10^0}$ at $\theta \ll 1$
and $\order{10^1}$ at $\theta \simeq 1$.
Hence, the percolation phase at $N_c =3$ is accessible at $T \ll \lqcd$, but at nuclear densities about 2-3 times that of the liquid phase,
which, at least using the scaling of Eq.~\ref{munclims}, are not yet confining.
This regime is somewhat lower in $T$ as that examined in e.g.~\cite{quarkyonic2,quarkyonic4,quarkyonic5}, making it likely that
percolation dynamics is relevant in proto-neutron stars (a similar regime to that examined in \cite{proto1,proto2})
rather than lower energy colliders \cite{low1,low2,low3,low4}.
In fact, given that at $N_f=N_c$  deconfinement and nuclear matter parametrically coincide, the crucial parameter determining the existence 
of a confined but percolating phase, rather than an ``early deconfinement'' at $\mu_Q \sim \lqcd$, might be the strange quark mass \cite{janstrange,jansbook}.    
A further uncertainity is the influence of the $\alpha=1$ component
(see Eq.~\ref{eq:b-to-b-probability} and following discussion).
If at $N_c \sim 3$ this component dominates, the percolation $N_c$-$\rho_B$ line is considerably less flat than shown in Fig.~\ref{percdecrhop} (note the different orientation of the $N_c$ axis in Fig. \ref{fig:samplecurves} and Fig. \ref{percdecrhop}) , while the deconfinement line is unaffected.
This has the effect of increasing the ``critical $N_c$'' where deconfinement and percolation cross; hence,
just as in the case of a ``light strange quark'' or in the presence of antibaryons,
a physical percolating but confined phase becomes less likely.

Given the quantitative roughness of the models considered here, however, these are in no way definite conclusions.
If anything, these results are much more encouraging for phenomenology than those of \cite{percolation1},
which suggested that at $N_c=3$ the percolating regime was strictly inaccessible.   
Given the uncertainties illustrated above, we will devote the next two sections, \ref{sectheo} and \ref{secexp}, to exploring some theoretical
and phenomenological aspects of the percolating phase, in view of both giving experimentalists and astrophysicists
some insight into how this phase could manifest, and sketching what the effective theory of this percolating phase might be like.

We close this section by comparing the percolating phase to the more ``usual'' nuclear matter in the large-$N_c$ limit.
One can ask how the phase considered here overlaps with the transition between the classical baryonic crystal considered in \cite{crystal} and normal nuclear matter.
The equivalence between the percolation transition and the onset of the classical regime for baryon dynamics can not be exact,
since percolation is insensitive to the number of flavors (as long as $N_f\geq1$, required for baryons to exist),
while the transition is driven not by $N_c/N_N \gg 1$ but rather $N_c/(N_f N_N) \gg 1$, where $N_N$ is the number of neighbors in a closely packed system \cite{vdw}.

However, the interplay between this transition and deconfinement has a similar, albeit weaker, dependence on $N_f$ ($\sim N_f^{-1/2}$ rather than $\sim N_f^{-1}$).
Since the critical $N_c$ for this transition is of $\order{10}$, the two transitions do approximately coincide for $N_f \sim \order{1}$.
This suggests that varying $N_c$ and $N_f$ separately could yield extremely non-trivial dynamics.
Such an ``experiment'', of course, is only possible on the lattice, perhaps by applying the strong-coupling methods of \cite{liqlat}
to the large-$N_c$ limit \cite{panero1,panero2}.

\section{Effective theory of the percolating phase}
\label{sectheo}
In this work we have used a simple but universal model, motivated by what we know about $N_c$-scaling
of the thermodynamics of Yang-Mills theories, to map the interplay of percolation and deconfinement across density,
temperature, and number of colors $N_c$.
We found a hitherto unexplored percolating phase, where confinement persists
(the Polyakov loop expectation value vanishes and baryons exist as semiclassical soliton states)
but quarks are able to propagate to arbitrarily high distances via inter-baryon tunneling,
whose probability is governed by a non-perturbative color-neutralizing propagator \`a la \cite{conf1,conf2,conf3}.   
Because of this, perturbative quarks and holes should be able to coexist at momenta $\sim \Lqcd$ in the background of baryonic ``classical'' potential wells.

\begin{figure}
\begin{center}
\includegraphics[width=10cm]{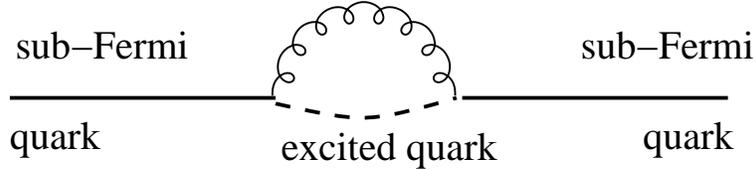}
       \caption{(color online)
A typical diagram introducing a temperature dependence on the pressure
}		
		\label{quarkdiagram}
\end{center}
\end{figure}

While, in many ways, our phase bears similarities to the ``quarkyonic phase'' conjectured in \cite{quarkyonic}
and explored in \cite{spiral1,spiral2,kojo}, there are also differences: in the quarkyonic phase, excitations about the Fermi surface
are assumed to be $\sim N_c^0$ at {\em all momenta}, and hence entropy continues to be $\sim N_c^0$, since sub-Fermi surface states carry no entropy.    
However the Gibbs-Duhem relation, linking energy density $e$ and conserved charge density $\rho_B$ to pressure $P$ and entropy density $s$,
\eq
	s = \frac{\de P}{\de T} = \frac{P+e-\mu_B \rho_B}{T}\;\;,
\qe
seems to demand a $s \sim N_c$ scaling in an interacting phase where $P \sim N_c$.
If the pressure scales as $P \sim N_c f(T)$, then the only way to avoid entropy density to scale as $\sim N_c$
is to have an equation of state strictly of the form
\eq
	P = N_c^0 f_1(\mu_B,T) + N_c^1 f_2(\mu_B)
\qe
({\em without} temperature dependence of sub-Fermi degrees of freedom).
At $T=0$ this is certainly the case, but quark-hole diagrams such as in Fig.~\ref{quarkdiagram} 
will inevitably add $\sim N_c f(\mu_B T)$ terms to the partition function, representing excited quark-hole states.
Exciting these states will cost momentum $\sim \rho_B^{-1/3}/N_c \sim N_c^{\alpha} \lqcd$ with $\alpha<1/2$,
hence is  not suppressed in the percolating phase.
Therefore, perturbations around the Fermi surface could still be colored, but confinement should be maintained at super-baryonic distances.
Provided we come up with a physical way to realize such a system, it is a reasonable way of identifying the percolation transition
demonstrated here with the phase conjectured in \cite{quarkyonic}.

\begin{figure}
\begin{center}
	\includegraphics[width=10.cm]{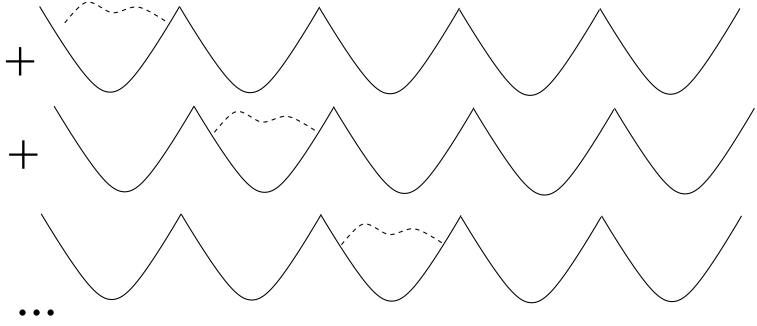}
	\caption{(color online) 
		An illustration of the ``free quark'' wavefunctions of the percolating phase.
		Baryons play the role of semiclassical potentials, analogously to atoms in a conductor.
		Quark wavefunctions are delocalized by tunnelling, and their Eigenstates are in superposition, analogously to electrons.
		Below percolation, tunnelling probability diverges for an infinitely large system, so quarks remain localized with an $\order{1}$ baryonic cluster.
	}
	\label{fig:chiral}
\end{center}
\end{figure}

A physical analogy arising from condensed matter physics is the metal-insulator 
transition:\footnote{Note that this analogy is not perfect, at least because the critical exponents of metal-insulator and percolation are different;
in percolation only the two-point correlation is relevant, while typically in metal-insulator transitions higher order correlation functions play a part.
The source terms in Eq.~\ref{wzwdef} could be used to represent such terms.}
the critical point of this transition can be understood as percolation of electrons across the semi-classical potential wells generated
by the atoms of the material \cite{metal}.
Tunneling probabilities, as well as electron-electron interactions, are what drives this transition.
The analogy with the picture presented here, with the baryons taking place of the atoms, is immediate.
In this picture, the low-$N_c$ confined system can only be an ``insulator'', with quarks of different hadrons not interacting or propagating.
At high $N_c$, however, a ``confined conductor'' phase is possible, in which the low-energy degrees of freedom are not $N_c^0$ heavy baryons,
but $\sim N_c$ quark-hole pairs:
the quarks and holes are confined in hadrons but, due to tunneling, can not be univocally assigned to a given hadron.
The free energy Eigenstates are therefore {\em superpositions} of quasiparticle quark wavefunctions across the whole system,
with characteristic momentum $k\sim \lqcd$, in the same way as the free energy Eigenstates of electrons in a metal are delocalized:
the ``free particle'' quark, hole ($q,h$) wavefunctions (combining color, flavor and spin) are not a continuum in $k$ but obey the Bloch constraint
\eq
	\label{psi}
	\Psi_{q,h}^k (x+\rho_B^{-1/3}) = \Psi_{q,h}^k(x) \exp\left[ i k \rho_B^{-1/3} \right]\;\;;
\qe
this forces the spectral function $\rho_q(k)$ to be of the form
\eq
	\label{rhoq}
	\rho_q(k) \simeq \sum_{n=0}^{\infty} \rho_q^n \left( n k \rho_B^{-1/3} \right)\;\;,
\qe
where $\rho_q^n(k)$ is a Lorentzian-type function.
Diagrammatically, this is shown in Fig.~\ref{fig:chiral}, where the solid lines represent the semiclassical ``mean field'' baryon potentials and the dashed lines 
the delocalized quark wavefunctions
Even if $N_c$ is ``high enough'' for pQCD quark-hole dynamics to be relevant,
this density of states is radically different from that of a free thermal quark-gluon plasma at high chemical potential,
where the spectral function is approximately constant, $\rho_q(k) \sim k^0$.

\begin{figure}
\begin{center}
	\includegraphics[width=10.cm]{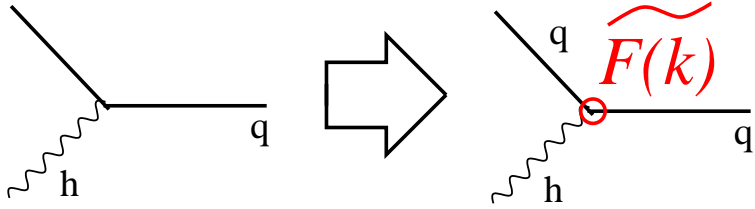}
	\caption{(color online)
		The difference between free-theory and quarkyonic-theory Feynman diagram expansions.
		The form factor $\widetilde{F}(k)$ is approximately the Fourier transform of the potential in Fig.~\ref{fig:chiral}.
	}
	\label{fig:formfactor}
\end{center}
\end{figure}

The combination between the asymptotically free nearly massless quarks with a spectral function inhomogeneous in momentum space such as Eq.~\ref{rhoq} is 
what ultimately enables the chiral inhomogeneities found in \cite{spiral1,spiral2}, and also in models such as \cite{chir1,chir2} and \cite{greiner1,greiner2}.
If such ``conductive'' quarks are in the asymptotic freedom regime in some limit, their dynamics can be computed perturbatively by adding 
form factors to quark propagators.
For scattering processes (such as quark-hole scattering of Fig.~\ref{figee} and the virtual excited quark of Fig.~\ref{quarkdiagram}) the quark 
and hole propagators will acquire form factors $\widetilde{F}(k)$ (Fig.~\ref{fig:formfactor})
\begin{equation}
\label{formfactor}
 \frac{k_0^{n-2}}{k^n} \rightarrow \frac{(\widetilde{F}(k))^2 k_0^{n-2}}{k^n}\;\;.
\end{equation}
$n=2$ would describe a 3D quasi-perturbative regime, while $n=4$ would be close to the Gribov limit described in section \ref{secmodel}.
The propagators of the non-abelian degrees of freedom in Eq.~\ref{wzwdef} would be similarly modified.
The form factor $\widetilde{F}(k)$ would be the Fourier transform of the lattice of nucleon mean fields shown in Fig.~\ref{fig:chiral} .
If quark wavefunctions are of the form of Eq.~\ref{psi}, then lower Fourier components $\Psi_{q,h}^k(x)$ can be color singlets
(note that the lowest mode is $\sim \rho_B^{1/3}$), while the higher modes are allowed to be colored, provided a compensating mechanism
(such as in \cite{neut1,neut2,neut3}) exists, neutralizing the color perturbations over scales larger than $\lqcde{-1}$.
Such color compensation must actually exist in the QGP as well to avoid paradoxes described, for example, in \cite{paradox};
however, such sub-$\lqcd$ correlations in the high-temperature regime would be negligible for any thermodynamic
property of the deconfined phase, since the microscopic scale of this system is above confinement $\sim 1/(N_c^2 N_f T) \ll \lqcde{-1}$.
This is not so obvious in a confined but ``quarkyonic'' phase;
we therefore must invent a way for $P,s \sim N_c$ to hold at $T\leq T_c$ and yet color neutrality be maintained at scales $\leq \lqcd$.

Condensed matter physics gives us another example of how this could work, namely spin-charge separation in 1D systems \cite{spincharge1,spincharge2}
(an effect that does indeed seem to be found in Non-abelian gauge theories \cite{gauge1,gauge2}):
in a 1D interacting fermion chain, spin and charge generally separate.
If, as suggested in \cite{spiral1,spiral2}, quarkyonic matter is governed by dimensional reduction,
such separation could provide the neutralizing force:
confinement would localize the color part of the wavefunction only, while allowing spin and charge to propagate as $N_c$ copies of a color-singlet field.   

While the quantitative development of such a theory is left for future work, we shall present a sketch of how this works.
We use the approach of \cite{spiral1}; the effective Lagrangian along an infinite percolating chain of quarks will reduce,
for $N_c$ colors and $N_f$ flavors, to
\eq
	S =  S_{2 N_f}[h_{\mathrm{color}}] + S_{N_c}[h_{\mathrm{flavor}}]\;\;.
	\label{wzw_action}
\qe
where $h_{\mathrm{flavor,color}}$ are separate ``flavor'' and ``color'' fields; note that each comes with ``redundant'' copies of the other.

Neglecting color neutralization, both $S_{2 Nf}$ and $S_{Nc}$ should have the following WZNW-inspired
form \cite{kojo,spiral1,wzw1,wzw2,wzw3,wzw4,wzw5,wzw6,wzw7,wzw8,wzw9}:
\eqa
	\label{wzwdef}
	S_{k}[\ell]
		&=& k\ {\mathrm tr} \bigg[
		\frac{1}{16\pi}
		\int d^2x \; f\left( x,\ell,\partial_\mu \ell \right) \partial_\mu \ell \partial^\mu \ell^{-1} \nonumber \\
	& & \qquad + \frac{1}{24\pi}
		\int d^3x \; \epsilon^{\mu \nu \lambda}
		(\ell^{-1} \partial_\mu \ell) (\ell^{-1}\partial_\nu \ell) (\ell^{-1} \partial_\lambda \ell)
		\bigg] + J_{branch}. 
\qea
The charges coming in and out of each effectively 1D chain through the branching of the percolation clusters can be represented by source terms $J_{branch}$.
At the percolation transition, they can be defined by the requirement of the 3D conformal invariance of the system.

If $f\left( x,\ell,\partial_\mu \ell \right)=1$, this Lagrangian reduces to \cite{wzw1,wzw2,wzw3,wzw4,wzw5,wzw6,wzw7,wzw8,wzw9},
and this will be approximately the case for the flavor part.
For the color part, however, \mbox{$f(\cdots) = g(x) h(\ell,\partial_\mu \ell)$} can be used as a mean field:
\begin{description}
	\item[baryons] can be represented by ``mean field wells'', having the form 
		\eq
			\ave{g} (x) = G_0 \left( 1 - \sum \hat{g}_n(x) e^{i  n \rho_B^{1/3}} \right)\;\;,
		\qe
		where $G_0 \sim N_c$.
		This forces any wavefunction for $\ell$ to be centered around Eq.~\ref{psi}.
	\item[neutralization] can be enforced by making  $f\left( x,\ell,\partial_\mu \ell \right)=1$
		trigger a large background field \cite{neut1} for color non-singlet states of momentum $p > \lqcd$.
		This effective mass could behave in a similar way as the color chemical potentials discussed in \cite{neut2,neut3}.
\end{description}
It can be seen that, with this Ansatz, $N_c$ ``flavor'' excitations of arbitrary frequency and
``color'' excitations of momentum $k \sim \lqcd$ survive.
For large $N_c$, these will dominate the entropy.

We thus recovered the premises of \cite{quarkyonic}, with equations such as \ref{eq:propagator_t} as a phenomenological form of the mean-field compensator $f(x,p)$.
We therefore arrive, {\em from the percolating side}, at a consistent physical justification of the Ans\"atze used in the first part of the paper.
When $N_c$ is below the percolation threshold, Eq.~\ref{psi} becomes unphysical,
because the probability of the wavefunction to tunnel more than one baryonic distance is vanishing.
Above the percolation threshold, quark wavefunctions are assigned not to a baryon,
but to {\em all} baryons in an infinite percolating chain following Eq.~\ref{psi}.
Quark-hole excitations will obey an effective action given by \ref{wzw_action}.
In thermal equilibrium, the flavor part of the wavefunction should yield entropy
and pressure $\sim N_c$ even if color is neutral at super-baryonic scales.

Why has nothing similar been observed in gauge/string duality, and how {\em can} this phenomenon be characterized in such a picture?
Since the percolation transition itself is driven by $N_c$, to model the percolation point one would have to include leading-order $g_s$ corrections,
where $g_s$ is the string coupling constant \cite{percolation1}.
The percolating phase, however, is in the {\em low}-$g_s$ limit, and therefore could in principle be seen by constructions
which include baryons in the semiclassical gravity limit, such as \cite{rozali,mateos}.
Yet nothing in these works suggests that the dense phase is anything different from a ``dense nuclear gas''.
Quark wavefunctions might be delocalized, yet this results in no additional degrees of freedom at the level of the entropy density and pressure.
 
The problem is that the argument in \cite{quarkyonic} assumes asymptotic freedom.
Even implementations such as Klebanov-Strassler \cite{ks,mia} do not have asymptotic freedom but rather
asymptotic $\mathcal{N}=4$ SYM with large $\lambda$ for ``hard'' momentum exchange.
If this transition will appear in gauge/string duality, it will be subleading in $\alpha'$.
The percolating regime occurs at low $g_{s}$ (high $N_c$) but higher string tension $\alpha'$ (lower $\lambda$),
while the ``nuclear matter'' phase discussed in \cite{lippert,rozali} happens in the weak limit of both $g_{s}$ and $\alpha'$.   

Typically, in gauge/string constructions \cite{lippert,rozali,mateos} baryons are represented by stacks of $D7$-branes,
with the ``nuclear matter'' phase being represented as a deformation of a string hanging from charged D-branes due to the charge on the brane \cite{lippert}.
The extra entropy scaling of quarkyonic matter must therefore be driven, in the gauge/gravity picture, by the appearance of KK modes in such a hanging string.
We conjecture, therefore, that quarkyonic matter of the type we discuss arises at low $g_s$ and moderate $\alpha'$.
Subleading corrections in $g_s$ will give rise to the $N_c$ percolation transition between percolating and non-percolating matter \cite{percolation1},
and perhaps the baryon quantum-to-classical transition discussed in \cite{vdw}.

\section{Phenomenology of the percolating phase}
\label{secexp}
Our calculations show that seeing this phase transition in future experiments in our $N_c=3$ world
\cite{low1,low2,low3,low4} {\em might} be possible, provided low-$T$, high-$\rho_B$ regions are accessible.
This makes it desirable to extend the above discussion and develop some phenomenology for the percolating phase.
As in Section \ref{sectheo}, the quantitative aspect of this is left for future work \cite{vogel}, but we can let the analogy with the metal-insulator transition guide us:
a universal characteristic signature of such a system is the appearance of band gaps in the spectral function of charge carriers due to
Eq.~\ref{rhoq} and the (weak) interactions between neighboring charge carriers \cite{gapref}.      This discussion is based on the assumption that baryon distributions in quarkyonic matter are more or less regular, allowing for regular band gamps in momentum space to form.   While this might not be unreasonable at high density \cite{maruhn}, as shown in \cite{vogel}, irregularities can introduce chaotic event-by-event fluctuations in electromagnetic form factors.   A more quantitative calculation is needed to assess the effect of these.

\begin{figure}[t]
\begin{center}
\includegraphics[width=15.cm]{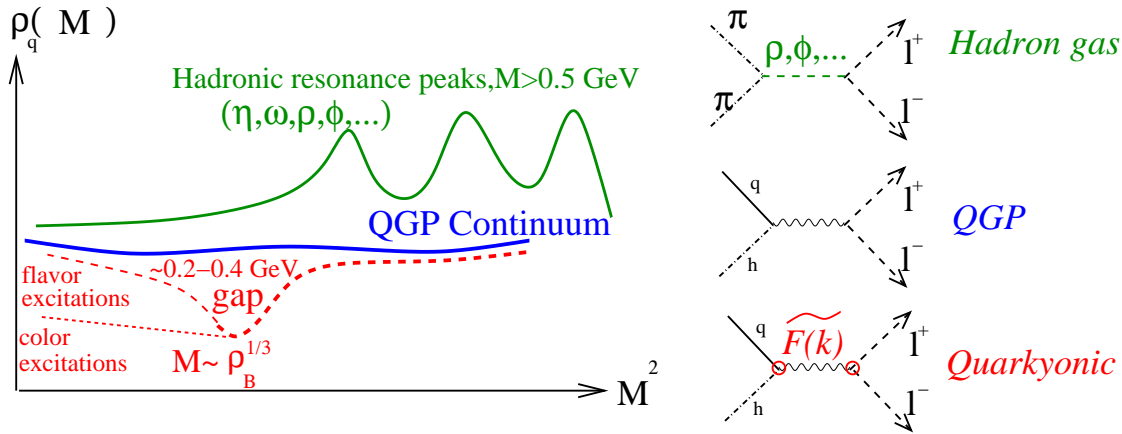}
\caption{
	(color online) Sketch of what the $ \ell^+ \ell^-$ spectral 
function 
could look like in percolating quarkyonic matter, in a QGP and in a resonance-dominated hadron gas.
		The gap in quarkyonic matter arises because no delocalized free quark states can exist around $k \sim \rho_B^{1/3}$ 
(the form factor in Eq.~\ref{formfactor} vanishes).
		At frequencies below the gap the color part of the wavefunction stops contributing, but the flavor part might still be present.
	}
	\label{figee}
\end{center}
\end{figure}

Such band gaps, of mass $\sim 250$-$400$ MeV (well below any resonance mass) could be directly detected in electromagnetic probes
(the spectral function of $e^+ e^-$ pairs in heavy ion collisions):
as the sketch in Fig.~\ref{figee} shows, common \mbox{$q \hspace{0.32em} \mathrm{hole} \to \gamma \to l^+ l^-$} scattering will give
an approximately flat spectral function for an unperturbed high-$\mu$ QGP.
If quark wavefunctions are delocalized across potential wells of size $\sim \lqcd$, $M^2 \sim \rho_B^{2/3}$ will be suppressed,
analogously to the scattering of x-rays by electrons in a conducting metal, due to the suppression of delocalized quark states around that frequency: 
the form factors of Eq.~\ref{formfactor} will be $\widetilde{F}(k = 1/\rho_B^{-1/3} ) \ll 1$, and that will depress the scattering cross-section shown in Fig.~\ref{figee}.
Below that frequency, color-neutralizing effects might suppress the color-part of the spectral function, but the flavor part of the spectral function can still contribute.    

Heavy-ion collisions at SPS and RHIC energies have yielded a continuum reminiscent of the QGP spectral function \cite{na61,phenix} on the top of peaks associated
to the decay of hadronic resonances ($\omega,\rho,\eta,\ldots$),
so perhaps the band gap structure can be searched for in upcoming lower energy experiments \cite{low1,low2,low3,low4}.

Alternatively, quarkyonic percolation as described here could be detected in the phenomenology of neutron stars and proto-neutron stars \cite{proto1,proto2}.
Quarkyonic matter would appear at a pressure about $\sim N_c=3$ times that of nuclear matter at the same density, temperature and chemical potential,
while maintaining a heat capacity and an energy density comparable to that of nuclear matter.
The extra boost in pressure is analogous to the way the electron gas dominates pressure in a metal.    
Such stiffer equations of state are desirable for stars such as 
\cite{naturestar}.\footnote{We would like to thank Irina Sagert for discussions regarding this topic.}

Furthermore, a quarkyonic phase in {\em proto-}neutron stars might be crucial in the dynamics of supernovae.
The effective stiffening of the equation of state might affect the early postbounce supernova dynamics and/or black hole formation
times during the core-collapse of massive stars.
The first effect is interesting in connection to the shock-stalling problem found in e.g.~\cite{noproto}. 
In \cite{proto1,proto2} this problem was solved with a more traditional deconfinement transition making the equation of state {\em softer},
since both equilibrium energy density and pressure increase at deconfinement, and the mixed phase drives the speed of sound to zero.    

The percolation transition looks significantly different in a way that 
might make it easier to maintain a shock-wave. In the $N_c \to \infty$ 
limit the phase transition line is vertical in the $T$-$\mu_B$ plane 
(regions I and II in Fig.~\ref{figphase}.  The real $N_c=3$ world has a 
\mbox{$\sim 30\%$} curvature correction), and hence both sides in the 
Clausius-Clayperon 
equation diverge. Percolation, however, implies a second order phase 
transition, hence the change in pressure ($P/(T\rho_B)$ jumps by $\sim 
N_c=3$ when $\rho_B$ crosses the percolation threshold, as quarks start exerting pressure) can only be {\em gradual}
with density (the ``jump'' is a rapid but smooth cross-over at any finite $N_c$), and 
there is no mixed phase, or jump in energy density or 
heat capacity ($e/(T \rho_B)$ stays approximately constant as $\rho_B$ is varied). 
It would be very interesting to assess the effect 
of an equation of state with such a transition 
 in calculations such as \cite{proto1,proto2,nakazato,hempel,prakash}.

In conclusion, we have studied the interplay between percolation and deconfinement in Yang-Mills matter at finite number of colors, temperature and density.
We find that these transitions exhibit a non-trivial dependence on $N_c$, suggesting that, at least for thermodynamics,
we can not automatically assume QCD is in the ``large-$N_c$ limit'' at $T,\mu_Q \sim \lqcd$.
Our calculations, however, show that the percolating phase {\em could} appear for $\rho_B \sim (0.125$ -- $3)\lqcde{3}$, provided quarks at this density are still confined.
Naive scaling in number of colors and flavors suggests they are, although we can not say this with certainty.
We have speculated what the dynamics of the percolating phase looks like and how it is related to popular approaches (such as the gauge/string duality)
for describing Yang-Mills matter in the same regime.
Even if the findings here will not be confirmed experimentally, characterizing them more rigorously on the lattice and in the
gauge/string correspondence opens quite a few questions, the solutions of which could help us clarifying the qualitative structure of Yang-Mills theories.
As a contact with our $N_c=3$ world can not be excluded, we have closed by suggesting experimental and astrophysical signatures of the percolating phase.

We thank Chris Hooley, Irina Sagert, Yonah Lemonick, Marco Panero, Larry McLerran, Pietro Giudice and Thomas Cohen for discussions.
We acknowledge the financial support received from the Helmholtz International
Centre for FAIR within the framework of the LOEWE program
(Landesoffensive zur Entwicklung Wissenschaftlich-\"Okonomischer
Exzellenz) launched by the State of Hesse.
GT also acknowledges support from DOE under Grant No. DE-FG02-93ER40764

\end{document}